\documentclass[prd,twocolumn,floatfix,amsmath,nofootinbib,amssymb,floatfix]{revtex4}
\usepackage{graphicx,color,dcolumn,booktabs,bm}
\usepackage{longtable,lscape}
\usepackage{pdfpages}
\usepackage{txfonts}
\usepackage{overpic}
\usepackage{amssymb}
\usepackage{makecell}
\usepackage{indentfirst}
\usepackage{feynmf}   
\usepackage{slashed}  
\usepackage{cases}
\usepackage{color}
\usepackage{multirow}
\usepackage{threeparttable}
\usepackage{epstopdf}
\usepackage{enumerate}
\usepackage{diagbox}
\usepackage{graphicx,color,dcolumn,booktabs,bm}
\usepackage[colorlinks,
            citecolor=blue,
            anchorcolor=red,
            menucolor=red,
            linkcolor=red,
            filecolor=red,
            runcolor=red,
            urlcolor=blue,
            frenchlinks=true]{hyperref}
\begin{document}
\title{Spectroscopic properties of $1F$-wave singly bottom baryons}
\author{Zi-Le Zhang$^{1,2,4}$}
\author{Si-Qiang Luo$^{1,2,3,4}$}\email{luosq15@lzu.edu.cn}

\affiliation{
$^1$School of Physical Science and Technology, Lanzhou University, Lanzhou 730000, China\\
$^2$Lanzhou Center for Theoretical Physics,
Key Laboratory of Theoretical Physics of Gansu Province,
Key Laboratory of Quantum Theory and Applications of MoE,
Gansu Provincial Research Center for Basic Disciplines of Quantum Physics, Lanzhou University, Lanzhou 730000, China\\
$^3$MoE Frontiers Science Center for Rare Isotopes, Lanzhou University, Lanzhou 730000, China\\
$^4$Research Center for Hadron and CSR Physics, Lanzhou University $\&$ Institute of Modern Physics of CAS, Lanzhou 730000, China
}

\begin{abstract}
This study investigates the mass spectra and decay behaviors of the experimentally unobserved $1F$-wave singly bottom baryons. Calculating their mass spectra could provide crucial guidance for determining their spectroscopic positions. Additionally, by analyzing their decay properties, we could predict the important decay channels, which are essential for experimental searches and quantum number assignments. Our calculations aim to support ongoing experimental and theoretical efforts in singly bottom baryon spectroscopy.
\end{abstract}
\maketitle

\section{Introduction}\label{sec:Introduction}
Significant advancements in science and technology have continuously propelled the development of hadron physics. In particular, the upgrades and improvements of large-scale hadron experiments~\cite{LHCb:2014set,CMS:2008xjf} such as LHCb and CMS have enabled measurements of hadron states with higher precision, higher energy ranges, and higher luminosity. This provides an excellent platform for the discovery of more hadrons and the study of their internal structures. Among these areas, hadron spectroscopy has consistently been a focal point of research. By studying hadron spectroscopy, we can explain experimentally observed hadron states, thereby aiding our understanding of the non-perturbative behavior of quantum chromodynamics (QCD) in the low energy region~\cite{Chen:2016spr,Guo:2017jvc,Cheng:2015iom,Brambilla:2019esw,Liu:2019zoy,Chen:2022asf,Meng:2022ozq,Dong:2021juy,Cheng:2021qpd}.

\begin{figure}[htbp]
\centering
\includegraphics[width=8.6cm]{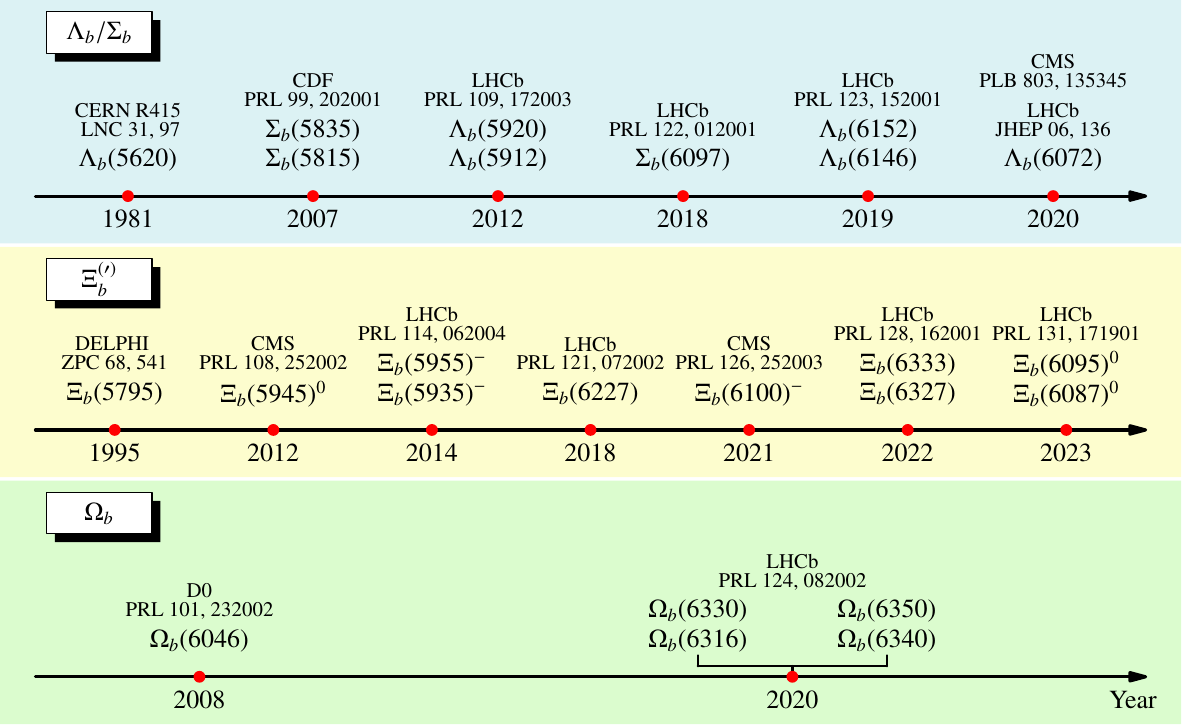}
\caption{The singly bottom baryons observed in experiments, with data sourced from Refs.\cite{DELPHI:1995jet,D0:2008sbw,LHCb:2019soc,CMS:2020zzv,Basile:1981wr,CDF:2007oeq,LHCb:2012kxf,LHCb:2018haf,LHCb:2020lzx,CMS:2012frl,LHCb:2014nae,LHCb:2018vuc,CMS:2021rvl,LHCb:2021ssn,LHCb:2023zpu,LHCb:2020tqd}.}
\label{fig:observation}
\end{figure}

Furthermore, predicting experimentally undiscovered hadron states and providing improved avenues for their experimental discovery, ultimately contributing to the enrichment of hadron families, is also a crucial aspect of current research. The study of singly bottom baryons constitutes a significant area within the hadron families~\cite{Klempt:2009pi,Gross:2022hyw,ParticleDataGroup:2024cfk}. To better illustrate the current state of observations concerning singly bottom baryons, we present Fig.~\ref{fig:observation} to outline the discovery times and experimental Collaborations for all known singly bottom baryons \cite{DELPHI:1995jet,Basile:1981wr,CDF:2007oeq,D0:2008sbw,CMS:2012frl,LHCb:2014nae,LHCb:2023zpu,LHCb:2018haf,LHCb:2018vuc,LHCb:2020tqd,CMS:2021rvl,CMS:2020zzv,LHCb:2012kxf,LHCb:2019soc,LHCb:2021ssn,LHCb:2020lzx}. As can be seen, over twenty singly bottom baryons have been observed, with the vast majority having been discovered in this century \cite{LHCb:2019soc,CMS:2020zzv,CDF:2007oeq,LHCb:2012kxf,LHCb:2018haf,LHCb:2020lzx,CMS:2012frl,LHCb:2014nae,LHCb:2018vuc,CMS:2021rvl,LHCb:2021ssn,LHCb:2023zpu,LHCb:2020tqd}. Consequently, the study of unobserved singly bottom baryon states through the lens of already discovered singly bottom baryons presents a promising opportunity.

If we ignore the isospin-multiples, except the $\Omega_b^*$ baryon, the remaining $1S$-wave singly bottom baryons have been observed~\cite{DELPHI:1995jet,Basile:1981wr,CDF:2007oeq,D0:2008sbw,CMS:2012frl,LHCb:2014nae}. In addition, during the past decades, a series of $1P$ candidates~\cite{LHCb:2012kxf,LHCb:2023zpu,LHCb:2018haf,LHCb:2018vuc,LHCb:2020tqd,CMS:2021rvl,CMS:2020zzv}, several $1D$ \cite{LHCb:2019soc,LHCb:2021ssn} and $2S$ \cite{LHCb:2020lzx} candidates are established. These observations greatly promote our understanding of singly bottom baryons. However, the $1F$-wave singly bottom baryons are still unobserved. Theorists have provided reasonable placements for these observed states within the singly bottom baryon family by analyzing their mass spectrum and decay properties~\cite{Korner:1994nh,Liu:2007fg,Karliner:2008sv,Aliev:2016xvq,Azizi:2020ljx,Wang:2020pri,Zhang:2008iz,Ebert:2005xj,Ghalenovi:2014swa,Thakkar:2016dna,Wei:2016jyk,Ghalenovi:2018fxh,Wang:2022dmw,Li:2023gbo,Weng:2024roa,Li:2024zze,Yang:2019cvw,Mao:2020jln,Chen:2020mpy,Liang:2020kvn,Yang:2020zrh,Majethiya:2011ry,Zhou:2023xxs,Chen:2019ywy,Aliev:2018vye,Aliev:2018lcs,Yang:2018lzg,Chen:2018orb,Cui:2019dzj,Liang:2020hbo,Wang:2022zcy,Bijker:2020tns,Agaev:2017ywp,Kakadiya:2021jtv,Chen:2021eyk,Tan:2023opd,Wang:2018fjm,Wang:2019uaj,Yao:2018jmc,Luo:2024jov}. The $1F$-wave singly bottom baryons, being located spectroscopically close to these already discovered singly bottom baryons~\cite{Ebert:2011kk,Yu:2022ymb,Li:2022xtj,Peng:2024pyl}, and the existence of these known states~\cite{ParticleDataGroup:2024cfk}, provide a sound basis for our study of the yet-unobserved $1F$-wave singly bottom baryons. Therefore, we conduct a spectroscopic analysis of the $1F$-wave singly bottom baryons to provide theoretical support for advancing their discovery.

Several theoretical models are employed to investigate the mass spectrum of singly bottom baryons, including the relativized quark model \cite{Capstick:1986ter}, the heavy-quark–light-diquark model \cite{Ebert:2011kk,Chen:2018vuc}, the Regge trajectory model \cite{Pan:2023hwt}, the Chiral quark model~\cite{Yang:2017qan}, and so on \cite{Garcilazo:2007eh,Mao:2015gya,Yang:2022oog,Wang:2024rai}. These models have achieved considerable success in describing baryon spectrum \cite{Capstick:1986ter,Ebert:2011kk,Chen:2018vuc,Yang:2017qan,Mao:2015gya,Garcilazo:2007eh,Yang:2022oog,Wang:2024rai,Xiao:2020oif,Liang:2019aag,Lu:2019rtg,Xiao:2020gjo,Chen:2014nyo,Pan:2023hwt}. In addition, the nonrelativistic potential model serves as a practical approach for handling singly heavy baryons~\cite{Chen:2016iyi,Luo:2023sne,Luo:2023sra,Peng:2024pyl}. In this work, we employ the nonrelativistic potential model in conjunction with the Gaussian expansion method (GEM) \cite{Hiyama:2003cu} to solve the three-body Schr\"odinger equation, thereby obtaining the mass spectrum of the $1F$-wave singly bottom baryons.

The mass spectrum alone provides limited information for the discovery of $1F$-wave singly bottom baryons. At this point, the decay behaviors of hadrons play a crucial role in further understanding their properties. Several theoretical studies have investigated the radiative decays of singly bottom baryons using different methods \cite{Ivanov:1999bk,Aliev:2009jt,Wang:2018cre,Aliev:2014bma,Yang:2019tst,Wang:2009cd,Ortiz-Pacheco:2023kjn,Wang:2010xfj,Hazra:2021lpa,Wang:2017kfr,Zhu:1998ih,Tawfiq:1999cf,Li:2017pxa,Jiang:2015xqa}, revealing different behaviors in the radiative decay widths for hadrons with the same isospin but different charge states. The Ref. \cite{Peng:2024pyl} further investigated the radiative decays of $1F$-wave singly bottom baryons. Moreover, many theoretical works have presented strong decay studies of singly bottom baryons in the $1S$, $1P$, $1D$, and other partial waves \cite{Majethiya:2011ry,Agaev:2017ywp,Aliev:2018vye,Aliev:2018lcs,Yang:2018lzg,Chen:2018orb,Cui:2019dzj,Liang:2020hbo,Wang:2022zcy,Zhou:2023xxs,Chen:2019ywy,Yang:2020zrh,Bijker:2020tns,Liang:2020kvn,Kakadiya:2021jtv,Chen:2021eyk,Tan:2023opd,Wang:2018fjm,Wang:2019uaj,Yao:2018jmc,Luo:2024jov}, but the investigations of $1F$-wave singly bottom baryons are relatively scarce. Compared to radiative decay studies, the strong decays of singly bottom baryons play a dominant role in their total width. Therefore, we further study the Okubo-Zweig-Iizuka (OZI) allowed two-body strong decays of the $1F$-wave singly bottom baryons using the quark pair creation (QPC) model~\cite{Micu:1968mk,LeYaouanc:1972vsx,LeYaouanc:1973ldf,Barnes:2007xu,Ackleh:1996yt}. By analyzing their partial and total decay widths, we identify the potentially dominant decay channels that could lead to their discovery.

This paper is organized as follows. In Sec. \ref{sec:mass}, we present the mass spectrum of singly bottom baryons using the nonrelativistic potential model and the Gaussian expansion method. In Sec. \ref{sec:Decays}, we analyze the OZI-allowed two-body strong decay widths of the $1F$-wave singly bottom baryons. In Sec. \ref{sec:Summary}, we conclude with a short summary.

\section{Mass spectrum calculations with nonrelativistic potential model}\label{sec:mass}
A singly bottom baryon consists of two light quarks $q_1$ and $q_2$ ($q = u, d, s$) and a heavy bottom quark $Q_3$ ($Q_3=b$). Owing to $SU(3)$ flavor symmetry, the flavor wave function of the singly bottom baryons can be decomposed into a $\bar{3}_f$ and a $6_f$ representation: $3_f \otimes 3_f = \bar{3}_f \oplus 6_f$. The anti-symmetric $\bar{3}_f$ representation includes the $\Lambda_b^0$, $\Xi_b^0$, and $\Xi_b^-$ baryons, while the symmetric $6_f$ representation includes the $\Sigma_b^{+}$, $\Sigma_b^0$, $\Sigma_b^-$, $\Xi_b^{\prime 0}$, $\Xi_b^{\prime -}$, and $\Omega_b^-$ baryons. The different symmetry of the flavor wave functions for the $\Lambda_b$ and $\Sigma_b$ baryons (and, similarly, the $\Xi_b$ and $\Xi_b^{\prime}$ baryons) leads to significant differences in their spectroscopic properties. The spatial structure of singly bottom baryons is often described using $\rho$ and $\lambda$ coordinates, where $\rho$ represents the relative position of the two light quarks, and $\lambda$ represents the position of the heavy quark relative to the center of mass of the two light quarks. Therefore, the basis for describing the singly bottom systems can be expressed as
\begin{equation}\label{eq:basis}
    \begin{split}
        |JM\rangle=|\phi^{{\rm color}}\phi^{{\rm flavor}}[[[s_{q_1}s_{q_2}]_{s_{\ell}}[n_\rho l_\rho, n_\lambda l_\lambda]_L]_{j_{\ell}}s_{Q_3}]_{JM}\rangle.
    \end{split}
\end{equation}
Here, $\phi^{\rm color}$ and $\phi^{\rm flavor}$ denote color and flavor wave function, respectively. $s_{q_1}$, $s_{q_2}$, and $s_{Q_3}$ stand for the spin quantum number of the quarks that make up the singly bottom baryon. $s_{\ell}$ and $j_{\ell}$ express the spin and total angular momentum of the light degree of freedom, respectively. The quantum numbers $n_{\lambda}$ and $l_{\lambda}$ correspond to the radial and orbital components of the $\lambda$-mode, while $n_{\rho}$ and $l_{\rho}$ represent the radial and orbital components of the $\rho$-mode. The $L$ denotes the total orbital angular momentum.

To accurately compute the mass spectrum of these singly bottom baryons, we employ a nonrelativistic potential model. The relevant Hamiltonian $\hat{H}$ is~\cite{Luo:2021dvj,Luo:2023sne,Luo:2023sra,Peng:2024pyl}
\begin{equation}\label{eq:H}
\hat{H}=\sum\limits_{i}\left(m_i+\frac{p_i^2}{2m_i}\right)+\sum\limits_{i<j}\left(H_{ij}^{\rm conf}+H_{ij}^{\rm hyp}+H_{ij}^{\rm so(cm)}+H_{ij}^{\rm so(tp)}\right),
\end{equation}
where $m_i$ and $p_i$ are the mass and momentum of the $i$-th constituent quark, respectively. The $H_{ij}^{\rm conf}$, $H_{ij}^{\rm hyp}$, $H_{ij}^{\rm so(cm)}$, and $H_{ij}^{\rm so(tp)}$ represent the spin-independent Cornell potential, the hyperfine spin-spin interaction, the color-magnetic piece of the spin-orbit term, and the Thomas-precession piece of the spin-orbit term, respectively. Their explicit expressions are 
\begin{equation}\label{eq:Vconf}
H_{ij}^{\rm conf}=-\frac{2}{3}\frac{\alpha_s}{r_{ij}}+\frac{1}{2}br_{ij}+\frac{1}{2}C,
\end{equation}
\begin{equation}\label{eq:Vhyp}
\begin{split}
H_{ij}^{\rm hyp}=&\frac{2\alpha_s}{3m_im_j}\left[\frac{8\pi}{3}\tilde{\delta}(r_{ij}){\bf s}_i\cdot{\bf s}_j+\frac{1}{r_{ij}^3}S({\bf r},{\bf s}_i,{\bf s}_j)\right],
\end{split}
\end{equation}
\begin{equation}\label{eq:Vsocm}
\begin{split}
H_{ij}^{{\rm so(cm)}}=&\frac{2\alpha_s}{3r_{ij}^3}\left(\frac{{\bf r}_{ij}\times{\bf p}_i\cdot{\bf s}_i}{m_i^2}-\frac{{\bf r}_{ij}\times{\bf p}_j\cdot{\bf s}_j}{m_j^2}\right.\\
&\left.-\frac{{\bf r}_{ij}\times{\bf p}_j\cdot{\bf s}_i-{\bf r}_{ij}\times{\bf p}_i\cdot{\bf s}_j}{m_im_j}\right),
\end{split}
\end{equation}
\begin{equation}\label{eq:Vsotp}
H_{ij}^{{\rm so(tp)}}=-\frac{1}{2r_{ij}}\frac{\partial H_{ij}^{\rm conf}}{\partial r_{ij}}\left(\frac{{\bf r}_{ij}\times{\bf p}_i\cdot{\bf s}_i}{m_i^2}-\frac{{\bf r}_{ij}\times{\bf p}_j\cdot{\bf s}_j}{m_j^2}\right).
\end{equation}
In Eqs.~(\ref{eq:Vconf})-(\ref{eq:Vsotp}), the $\alpha_s$ represents the coupling constant of one-gluon exchange, $b$ is the strength of the linear confinement potential, and $C$ is a mass-renormalization constant. The
\begin{equation}
\tilde{\delta}(r)=\frac{\sigma^3}{\pi^{3/2}}{\rm e}^{-\sigma^2r^2}
\end{equation}
is the smearing function, and the $\sigma$ is a smearing parameter. The $S({\bf r},{\bf s}_i,{\bf s}_j)$ is the tensor operator, which is defined as 
\begin{equation}
S({\bf r},{\bf s}_i,{\bf s}_j)=\frac{3{\bf s}_i\cdot{\bf r}_{ij}{\bf s}_j\cdot{\bf r}_{ij}}{r_{ij}^2}-{\bf s}_i\cdot{\bf s}_j.
\end{equation}

Using the above Hamiltonian $\hat{H}$, we could solve the three-body Schr\"odinger equation via the GEM~\cite{Hiyama:2003cu} to obtain the masses and corresponding spatial wave functions of singly bottom baryons. The Gaussian wave function is written as
\begin{equation}
\begin{split}
\phi_{nlm}(\boldsymbol{r})=&N_{nl}r^l{\rm e}^{-\nu_nr^2}Y_{lm}(\hat{\boldsymbol{r}}),~~~N_{nl}=\sqrt{\frac{2^{l+2}(2\nu_n)^{l+\frac{3}{2}}}{\sqrt{\pi}(2l+1)!!}},
\end{split}
\end{equation}
where $\nu_n$ is the Gaussian range, i.e.,
\begin{equation}
\nu_n=\frac{1}{r_n^2},\;r_n=r_1a^{n-1}\;(n=1,2\cdots N_{\rm max}).
\end{equation}

After the introduction of the Gaussian basis, the complete wave function of the singly bottom baryons could be obtained by
\begin{equation}
|\psi_{JM}^{n_\rho^g n_\lambda^g}\rangle=|\phi^{{\rm color}}\phi^{{\rm flavor}}[[[s_{q_1}s_{q_2}]_{s_{\ell}}[\phi_{n_\rho^g l_\rho}({\bm \rho}) \phi_{n_\lambda^g l_\lambda}({\bm \lambda})]_L]_{j_{\ell}}s_{Q_3}]_{JM}\rangle,
\end{equation}
\begin{equation}\label{eq:PsiJM}
\begin{split}
|\Psi_{JM}\rangle=\sum\limits_{n_\rho^g n_\lambda^g}&C_{n_\rho^g n_\lambda^g}|\psi_{JM}^{n_\rho^g n_\lambda^g}\rangle.
\end{split}
\end{equation}
For convenience, we ignore the third component of the angular momentum in the coupling scheme. The $C_{n_\rho^g n_\lambda^g}$ in Eq.~(\ref{eq:PsiJM}) are obtained by the Rayleigh—Ritz variational method. In this scheme, the three-body Schr\"odinger equation could be solved with the following eigenvalue form
\begin{equation}\label{eq:EC}
\sum\limits_{n_\rho^g n_\lambda^g}H_{n^{g\prime}_\rho n^{g\prime}_\lambda n_\rho^g n_\lambda^g}C_{n_\rho^g n_\lambda^g}=E\sum\limits_{n_\rho^g n_\lambda^g}N_{n^{g\prime}_\rho n^{g\prime}_\lambda n_\rho^g n_\lambda^g}C_{n_\rho^g n_\lambda^g},
\end{equation}
where the $E$ and $C_{n_{\rho}^gn_{\lambda}^g}$ are the eigenvalues and eigenvectors. The Hamiltonian matrix elements and Gaussian basis overlaps are defined by
\begin{equation}
\begin{split}
H_{n^{g\prime}_\rho n^{g\prime}_\lambda n_\rho^g n_\lambda^g}=&\langle\psi_{JM}^{n_\rho^{g \prime} n_\lambda^{g \prime}}|\hat{H}|\psi_{JM}^{n_\rho^g n_\lambda^g}\rangle,\\
N_{n^{g\prime}_\rho n^{g\prime}_\lambda n_\rho^g n_\lambda^g}=&\langle\psi_{JM}^{n_\rho^{g \prime} n_\lambda^{g \prime}}|\psi_{JM}^{n_\rho^g n_\lambda^g}\rangle.
\end{split}
\end{equation}
Here, we should emphasize that $n_\rho^g$ and $n_\lambda^g$ are not the radial quantum numbers $n_\rho$ and $n_\lambda$ in Eq.~(\ref{eq:basis}). The $n_\rho^g$ and $n_\lambda^g$ could be extracted by sorting the eigenvalues and analysing the eigenvectors, and the procedure is similar to the solutions of the hydrogen atom and harmonic oscillator Schr\"odinger equations~\cite{Hiyama:2003cu}.

In calculations, $\{r_1,r_{\rm max},N_{\rm max}\}=\{\rho_1,\rho_{\rm max},N_{\rm max}^{\rho}\}$ and $\{r_1,r_{\rm max},N_{\rm max}\}=\{\lambda_1,\lambda_{\rm max},N_{\rm max}^{\lambda}\}$ are Gaussian parameters of the $\rho$- and $\lambda$-modes. We need to adjust these parameters to obtain the optimal solution. Meanwhile, the corresponding eigenvalues and eigenvectors can be obtained by changing the respective quantum numbers $s_\ell$, $l_\rho$, $l_\lambda$, $L$, etc.

\begin{table}[htbp]
\caption{Values of the parameters for the singly bottom baryons in the nonrelativistic potential model.}
\label{tab:parameter}
\renewcommand\arraystretch{1.25}
\begin{tabular*}{86mm}{@{\extracolsep{\fill}}ccccc}
\toprule[1.00pt]
\toprule[1.00pt]
System               &$\alpha_s$ &$b$ (GeV$^2$) &$\sigma$ (GeV) &$C$ (GeV)  \\
\midrule[0.75pt]
$\Lambda_b/\Sigma_b$ &0.560      &0.113         &1.600          &$-$0.573   \\
$\Xi_b^{(\prime)}$   &0.560      &0.126         &1.600          &$-$0.619   \\
$\Omega_b$           &0.578      &0.126         &1.732          &$-$0.597   \\
meson                &0.578      &0.144         &1.028          &$-$0.685   \\
\midrule[0.75pt]
\multicolumn{5}{c}{\mbox{$m_{u/d}=0.370~{\rm GeV}~m_{s}=0.600~{\rm GeV}~m_{b}=5.210~{\rm GeV}$}}\\
\bottomrule[1.00pt]
\bottomrule[1.00pt]
\end{tabular*}
\end{table}

\begin{figure*}
    \centering
    \includegraphics[width=\textwidth]{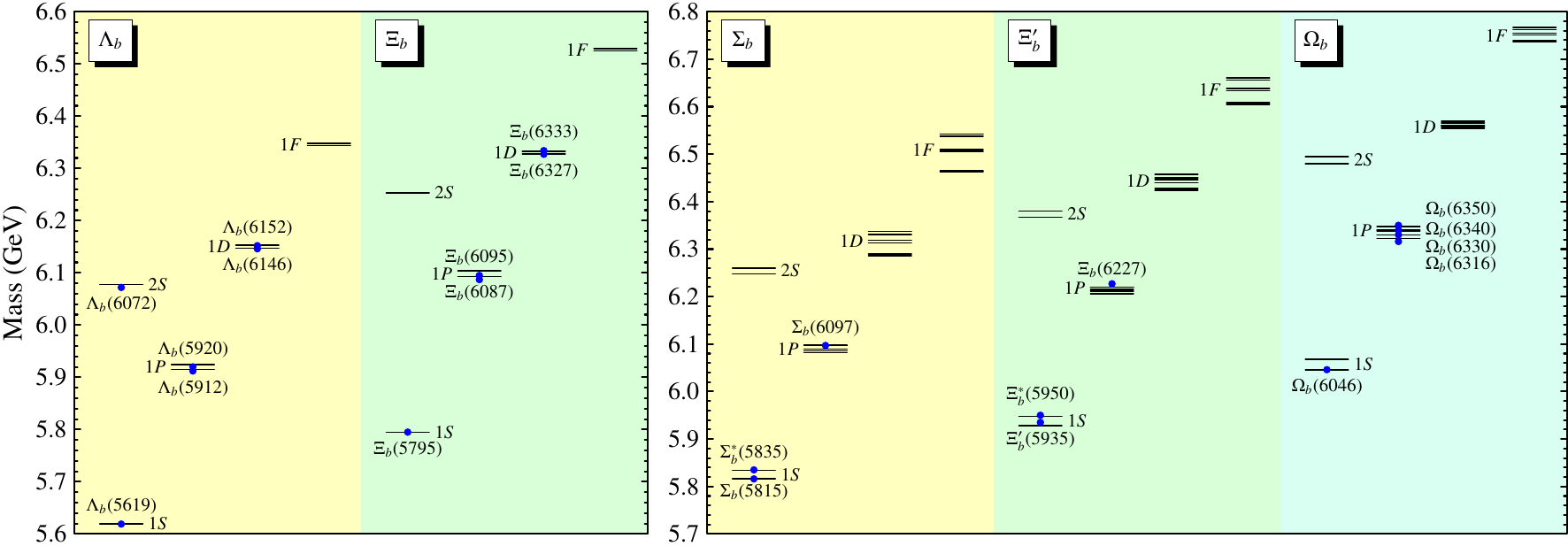}
    \caption{Comparison of our results and experimental measurements. The black solid short lines and blue dots represent our calculated results and the observed values from Particle Data Group (PDG) \cite{ParticleDataGroup:2024cfk}, respectively.}
    \label{fig:spectrum}
\end{figure*}

The parameters in the nonrelativistic potential model could be defined by the well-established singly bottom baryons from experiments \cite{ParticleDataGroup:2024cfk}, which are summarized in Table \ref{tab:parameter}. In Fig.~\ref{fig:spectrum}, we present our calculated masses of singly bottom baryons and make a comparison with the experimental data \cite{ParticleDataGroup:2024cfk}, which clearly shows that our results are in precise agreement with the experimental observations for the $1S$, $1P$, $1D$, and $2S$ states. The results indicate that the $\lambda$ single-mode excitation plays a significant role in the spectroscopy of currently observed singly bottom baryons. Similar results can be found in the Refs. \cite{Yoshida:2015tia,Pan:2023hwt,Peng:2024pyl}. Experimental data on the mass spectrum of the $1F$-wave singly bottom baryons are currently lacking. Therefore, investigating their mass spectrum is essential to determine their spectroscopic positions within the singly bottom baryon family.

Building on these preliminary results, we further calculated the masses of the $1F$-wave singly bottom baryons. To refer to the details of $1F$-wave singly bottom baryons, we present a compilation of our calculated results in Table \ref{tab:massspectrum}, including the assigned quantum numbers for each state and a comparison with other theoretical predictions. Similar results can also be obtained in Refs. \cite{Yoshida:2015tia,Pan:2023hwt,Peng:2024pyl}.

\begin{table*}
\centering
\caption{A comparison of our calculated masses for $\lambda$-mode excited $F$-wave singly bottom baryons with those from other theoretical articles \cite{Ebert:2011kk,Yu:2022ymb,Li:2022xtj}, with the listed masses in units of MeV. Here, the quantum numbers $s_\ell$, $l_\rho$, $l_\lambda$, $L$, $j_\ell$, and $J$ are defined consistently with Eq. (\ref{eq:basis}).}
\label{tab:massspectrum}
\renewcommand\arraystretch{1.15}
\begin{tabular*}{178mm}{@{\extracolsep{\fill}}cccccccccc}
\toprule[1.00pt]
\toprule[1.00pt]
		States  &$s_{\ell}$&$l_{\rho}$&$l_{\lambda}$&$L$&$j_{\ell}$&$J$                &This work  &Ref.~\cite{Ebert:2011kk} &Refs.~\cite{Yu:2022ymb,Li:2022xtj}\\
\midrule[0.75pt]
$\Lambda_b(1F,5/2^-)$ &0&0&3&3&3&$5/2$&6344 &6408                &6338                  \\
$\Lambda_b(1F,7/2^-)$ &0&0&3&3&3&$7/2$&6348 &6411                &6343                  \\
$\Xi_b(1F,5/2^-)$   &0&0&3&3&3&$5/2$ &6525 &6577                     &6518\\
$\Xi_b(1F,7/2^-)$   &0&0&3&3&3&$7/2$&6529 &6581                      &6523\\
$\Sigma_{b2}(1F,3/2^-)$ &1&0&3&3&2&$3/2$&6538 &6550                     &6538                   \\
$\Sigma_{b2}(1F,5/2^-)$ &1&0&3&3&2&$5/2$&6542 &6564                     &6542                  \\
$\Sigma_{b3}(1F,5/2^-)$ &1&0&3&3&3&$5/2$&6506 &6501                     &6538                    \\
$\Sigma_{b3}(1F,7/2^-)$ &1&0&3&3&3&$7/2$&6510 &6500                     &6542                   \\ 
$\Sigma_{b4}(1F,7/2^-)$ &1&0&3&3&4&$7/2$&6463 &6472                     &6538                  \\
$\Sigma_{b4}(1F,9/2^-)$ &1&0&3&3&4&$9/2$&6465 &6459                     &6542                   \\
$\Xi_{b2}^\prime(1F,3/2^-)$ &1&0&3&3&2&$3/2$&6656 &6675                     &6657\\
$\Xi_{b2}^\prime(1F,5/2^-)$ &1&0&3&3&2&$5/2$&6661 &6686                     &6660\\
$\Xi_{b3}^\prime(1F,5/2^-)$ &1&0&3&3&3&$5/2$&6634 &6640                     &6657\\
$\Xi_{b3}^\prime(1F,7/2^-)$ &1&0&3&3&3&$7/2$&6638 &6641                     &6660\\
$\Xi_{b4}^\prime(1F,7/2^-)$ &1&0&3&3&4&$7/2$&6605 &6619                     &6657\\
$\Xi_{b4}^\prime(1F,9/2^-)$ &1&0&3&3&4&$9/2$&6608 &6610                     &6661\\
$\Omega_{b2}(1F,3/2^-)$ &1&0&3&3&2&$3/2$&6762 &6763                     &6751\\
$\Omega_{b2}(1F,5/2^-)$ &1&0&3&3&2&$5/2$&6767 &6771                     &6754\\
$\Omega_{b3}(1F,5/2^-)$ &1&0&3&3&3&$5/2$&6751 &6737                     &6751\\
$\Omega_{b3}(1F,7/2^-)$ &1&0&3&3&3&$7/2$&6755 &6736                     &6754\\
$\Omega_{b4}(1F,7/2^-)$ &1&0&3&3&4&$7/2$&6736 &6719                     &6750\\
$\Omega_{b4}(1F,9/2^-)$ &1&0&3&3&4&$9/2$&6739 &6713                     &6754\\
		\bottomrule[1.00pt]
		\bottomrule[1.00pt]
	\end{tabular*}
\end{table*}

\section{OZI-Allowed two-body strong decay behaviors with QPC model}\label{sec:Decays}
In addition to the mass spectrum, analyzing the two-body strong decay of hadrons is essential for their discovery and for determining their quantum numbers. The QPC model~\cite{Micu:1968mk} is a powerful method for handling the OZI-allowed two-body strong decays and is widely utilized in the study of singly heavy baryons~\cite{Chen:2016iyi,Luo:2023sne,Luo:2023sra,Lu:2019rtg}. In this section, we employ the QPC model to investigate the partial and total decay widths of the $\lambda$-mode excited $1F$-wave singly bottom baryons.

\begin{table*}
\centering
\caption{The $\beta$ values of the hadrons associated with this work. The $\beta$ are given in units of GeV.}
\label{tab:beta}
\renewcommand\arraystretch{1.25}
\begin{tabular*}{178mm}{@{\extracolsep{\fill}}ccccccccc}
\toprule[1.00pt]
\toprule[1.00pt]
States&$\beta_\rho$ &$\beta_\lambda$ &States &$\beta_\rho$ &$\beta_\lambda$&States&$\beta_\rho$ &$\beta_\lambda$\\
\midrule[0.75pt]
$\Lambda_b(1S)$  &0.286 &0.365 &$\Xi_b(1S)$&0.297 &0.412&$\Omega_b(1S)$&0.279&0.440\\
$\Lambda_b(2S)$  &0.248 &0.194 &$\Xi_b(2S)$&0.255 &0.221&$\Omega_b^*(1S)$&0.274&0.426  \\
$\Lambda_b(1P)$  &0.262 &0.247 &$\Xi_b(1P)$&0.272 &0.278&$\Omega_b(2S)$&0.233&0.238 \\
$\Lambda_b(2P)$  &0.250 &0.155 &$\Xi_b(2P)$&0.258 &0.183&$\Omega_b^*(2S)$&0.234&0.233  \\
$\Lambda_b(1D)$  &0.251 &0.190 &$\Xi_b(1D)$&0.260 &0.213&$\Omega_b(1P)$&0.252&0.295  \\
$\Lambda_b(1F)$  &0.247 &0.160 &$\Xi_b(1F)$&0.254 &0.176&$\Omega_b(1D)$&0.243&0.230  \\
$\Sigma_b(1S)$   &0.215 &0.344 &$\Xi_b^\prime(1S)$ &0.245 &0.398&$\Omega_b(1F)$&0.236&0.188  \\
$\Sigma_b^*(1S)$ &0.212 &0.336 &$\Xi_b^*(1S)$&0.241 &0.386&$N$&0.280&0.324\\
$\Sigma_b(2S)$   &0.189 &0.189 &$\Xi_b^\prime(2S)$ &0.213 &0.219&$\Delta$&0.249&0.288\\
$\Sigma_b^*(2S)$ &0.190 &0.186 &$\Xi_b^*(2S)$      &0.214 &0.215&$\Lambda$&0.281&0.285\\
$\Sigma_b(1P)$   &0.197 &0.236 &$\Xi_b^\prime(1P)$ &0.223 &0.267&$\Sigma$&0.223&0.301\\
$\Sigma_b(1D)$   &0.191 &0.187 &$\Xi_b^\prime(1D)$ &0.217 &0.212&$\Sigma^*$&0.206&0.262\\
$\Sigma_b(1F)$   &0.187 &0.159 &$\Xi_b^\prime(1F)$ &0.212 &0.176&$\Xi$&0.287&0.317 \\
&$\beta_{\pi}=0.409$&$\beta_K=0.385$&$\beta_{K^*}=0.259$&$\beta_B=0.349$&$\beta_{B^*}=0.330$&$\Xi^*$&0.258&0.265\\
\bottomrule[1.00pt]
\bottomrule[1.00pt]
\end{tabular*}
\end{table*}

The transition operator of the QPC model reads as
\begin{equation}\label{eq:Troperator}
\begin{split}
\hat{\cal T}=&-3\gamma\sum_{m}\langle 1,m;1,-m|0,0\rangle\int{\rm d}^3{\bf p}_i\;{\rm d}^3{\bf p}_j\;\delta({\bf p}_i+{\bf p}_j)\\
          &\times\mathcal{Y}_1^m\left(\frac{{\bf p}_i-{\bf p}_j}{2}\right)\omega_0^{(i,j)}\phi_0^{(i,j)}\chi_{1,-m}^{(i,j)}b^\dagger_i({\bf p}_i)d^\dagger_j({\bf p}_j),
\end{split}
\end{equation}
where $\omega$, $\phi$, $\chi$, and $\mathcal{Y}$ are the color, flavor, spin, and spatial functions for the quark pair created from the vacuum, respectively. The $b_i^{\dagger}$ and $d_j^{\dagger}$ denote the creation operators of quark and antiquark. The dimensionless parameter $\gamma$ reflects the strength of quark-antiquark pair creation from the vacuum, which can be determined by fitting the width of the well-established states.

For an OZI-allowed process $A \to B + C$ with relative orbital angular momentum $L_{BC}$ and relative spin $S_{BC}$ between $B$ and $C$, the corresponding partial wave amplitude can be obtained by
\begin{equation}\label{eq:MAtoBC}
{\cal M}_{A\to BC}^{L_{BC}S_{BC}}(p)=\langle BC,L_{BC},S_{BC},p|\hat{\cal T}|A\rangle,
\end{equation}
where $p$ is the momentum of the particle $B$ in the rest frame of $A$. The partial width can be written as
\begin{equation}
\Gamma_{A\to BC}=2\pi\frac{E_BE_C}{M_A}p\sum\limits_{L_{BC}S_{BC}}|{\cal M}_{A\to BC}^{L_{BC}S_{BC}}(p)|^2,
\end{equation}
where the energies of the baryon $B$ and the meson $C$ are given by $E_B = \sqrt{M_B^2 + p^2}$ and $E_C = \sqrt{M_C^2 + p^2}$, respectively.

Inspired by research on meson decays \cite{Ackleh:1996yt,Close:2005se,Barnes:2007xu,Song:2015nia}, we adopt the simple harmonic oscillator (SHO) wave function to describe the spatial structure of hadrons. The expressions of the SHO wave function in coordinate and momentum spaces are
\begin{equation}\label{eq:shor}
R_{nlm}^{\rm SHO}(\beta,{\bm r})=\beta^{l+\frac{3}{2}}\sqrt{\frac{2n!}{\Gamma(n+l+\frac{3}{2})}}L_{n}^{l+\frac{1}{2}}(\beta^2r^2){\rm e}^{-\frac{\beta^2 r^2}{2}}r^l Y_{lm}(\hat{\bm r}),
\end{equation}
\begin{equation}\label{eq:shop}
\begin{split}
R_{nlm}^{\rm SHO}(\beta,{\bm P})=&\frac{(-1)^n(-{\mathrm i})^l}{\beta^{\frac{3}{2}+l}}\sqrt{\frac{2n!}{\Gamma(n+l+\frac{3}{2})}}L_{n}^{l+\frac{1}{2}}({P^2}/{\beta^2})\\
&\times{\mathrm e}^{-\frac{{P}^2}{2\beta^2}}P^lY_{lm}(\hat{\bm P}).
\end{split}
\end{equation}
Here, the $n$, $l$, and $m$ denote the radial, orbital, and magnetic quantum numbers, respectively. The $\beta$ is a parameter for scaling the SHO wave function.

Since a singly bottom baryon involves two spatial degrees of freedom, i.e., $\rho$- and $\lambda$-modes, two SHO wave functions are introduced to reduce the spatial structures of the singly bottom baryons. In this way, we need $\beta_\rho$ and $\beta_\lambda$ to depict the SHO wave functions for $\rho$- and $\lambda$-modes, respectively. The $\beta_{\bf \rho}$ and $\beta_{\bf \lambda}$ can be extracted with
\begin{equation}
\begin{split}\label{eq:beta}
\frac{1}{\beta_\rho^2}=&\int|\phi^{\rm spatial}({\bm \rho},{\bm \lambda})|^2{\bm \rho}^2{\rm d}^3{\bm \rho}{\rm d}^3{\bm \lambda},\\
\frac{1}{\beta_\lambda^2}=&\int|\phi^{\rm spatial}({\bm \rho},{\bm \lambda})|^2{\bm \lambda}^2{\rm d}^3{\bm \rho}{\rm d}^3{\bm \lambda},
\end{split}
\end{equation}
where
\begin{equation}
\phi^{\rm spatial}({\bm \rho},{\bm \lambda})=\sum\limits_{n_\rho^g n_\lambda^g}C_{n_\rho^g n_\lambda^g}[\phi_{n_\rho^g l_\rho}({\bm \rho}) \phi_{n_\lambda^g l_\lambda}({\bm \lambda})]_{LM_L}
\end{equation}
is the spatial wave function obtained from the potential model solved by GEM. Here, the $C_{n_\rho^g n_\lambda^g}$ are eigenvectors of Eq.~(\ref{eq:EC}). In this way, we have the following approximation
\begin{equation}
\phi^{\rm spatial}({\bm \rho},{\bm \lambda})\approx[R_{n_\rho l_\rho}^{\rm SHO}(\beta_\rho,{\bm \rho})R_{n_\lambda l_\lambda}^{\rm SHO}(\beta_\lambda,{\bm \lambda})]_{LM_L}.
\end{equation}
The formula for calculating the $\beta$ value, as given in Eq. (\ref{eq:beta}), provides an efficient and widely adopted method for studying the decay widths of singly heavy baryons~\cite{Chen:2016iyi,Luo:2023sne,Luo:2023sra,Peng:2024pyl}. The calculated $\beta$ values are presented in Table \ref{tab:beta}.

The wavefunction of $\pi$, $K$, and some other mesons involved in our work could be written as
\begin{equation}
    \begin{split}
        |\phi^{\rm color}\phi^{\rm flavor}[\phi_{NL}({\bf r})[s_{q_1}s_{q_2}]_S]_{JM}\rangle,
    \end{split}
\end{equation}
where the radial wavefunction $\phi_{NL}({\bf r})$ is taken from SHO wave function, which can be obtained through a similar approach as described in Eqs.~(\ref{eq:shor})-(\ref{eq:beta}). $s_{q_1}$ and $s_{q_2}$ stand for the spin quantum number of the quarks that make up these mesons. $S$ and $L$ denote the total spin quantum number and the total orbital angular momentum, respectively.

In addition, determining the dimensionless parameter $\gamma$ in Eq. (\ref{eq:Troperator}) is crucial for calculating the decay widths of the $1F$-wave singly bottom baryons. We determine $\gamma = 9.61$ by fitting to the experimental width~\cite{ParticleDataGroup:2024cfk} of the $\Sigma_b^*(5835)$ state. The value of $\gamma$ is employed in the global calculations in this work.

With these preparations, we now calculate the decay widths of these $1F$-wave $\Lambda_b$, $\Xi_b$, $\Sigma_b$, $\Xi^{\prime}_b$, and $\Omega_b$ baryons and analyze the results.

\subsection{$\Lambda_b(1F)$ baryons}
The $\Lambda_b(1S)$, $\Lambda_b(1P)$, $\Lambda_b(1D)$, and $\Lambda_b(2S)$ states are well-classified \cite{Basile:1981wr,LHCb:2012kxf,LHCb:2019soc,LHCb:2020lzx}, while the $\Lambda_b(1F)$ states remain experimentally elusive. Ongoing efforts aim to uncover them and achieve a more complete understanding of the $\Lambda_b$ family. The spectrum behavior provides essential insights for guiding the experimental search for these states. We now analyze their decay modes to further facilitate their identification.

Table \ref{tab:Lambdab1F} presents the calculated partial and total decay widths for the $\Lambda_b(1F)$ states. According to our calculations, the total width of the $\Lambda_b(1F, 5/2^-)$ state is approximately 59.4 MeV, with the corresponding branching fractions for the decay channels $N\bar{B}$, $N\bar{B}^*$, and $\Sigma_{b2}(1P, 3/2^-)\pi$ being 41.9$\%$, 30.0$\%$, and 19.7$\%$, respectively. For the $\Lambda_b(1F, 7/2^-)$ state, the total width is approximately 64.0 MeV, with the $N\bar{B}$, $N\bar{B}^*$, and $\Sigma_{b2}(1P, 5/2)\pi$ channels contributing 9.2$\%$, 68.6$\%$, and 19.2$\%$, respectively.

The results reveal that the $N\bar{B}$ and $N\bar{B}^*$ channels show larger branching ratios. The reason can be attributed to the fact that, under heavy quark symmetry, the $\Lambda_b(1F) \to N\bar{B}^{(*)}$ process preserves the $\rho$-mode reasonably well, which is a favorable decay mode. Because the branching ratios of the $\Lambda_b(1F,J^P)\to \Sigma_{b2}(1P,(J-1)^P)\pi$ processes are mainly contributions from $P$-wave \footnote{We should emphasize that ``$S$-", ``$P$-", ``$D$-", ``$F$-", etc. wave have two meanings. One is that the value of $l_\rho+l_\lambda$ for baryons, which is an inner quantum number. For example, we use $l_\rho+l_\lambda =3$ as $F$-wave, $l_\rho+l_\lambda =2$ as $D$-wave, and so on. The other is the relative orbital angular momentum of the final $BC$, i.e., $L_{BC}$ in Eq.~(\ref{eq:MAtoBC}). For example, in the $\Sigma_b^{*}\to \Lambda_b\pi$, $L_{\Lambda_b\pi}=1$, i.e., $P$-wave.} decays, with a smaller contribution from higher partial waves, the width of these processes is also relatively large. However, the other involving $\Sigma_b(1P)\pi$ channels can only proceed through particularly high partial waves, which results in their branching ratios being relatively small. In addition, the $\Lambda_b(1F) \to \Sigma_b(1S)\pi$ processes proceed through $D$-waves and even higher partial waves, with an exceptionally small width for these processes. Consequently, experimental investigations aimed at observing these $\Lambda_b(1F)$ states should prioritize the $N\bar{B}^*$ and $\Sigma_{b}(1P)\pi$ channels. Moreover, detailed studies of the $N\bar{B}$ channel are expected to provide valuable information regarding the characteristics of the $\Lambda_b(1F, 5/2^-)$ state.

\begin{table}
\centering
\caption{The calculated partial and total decay widths (in units of MeV) of the $\Lambda_b(1F)$ states. $M_f$ denotes the mass of the outgoing singly bottom baryons. The ellipsis ``$\cdots$" denotes decay channels with extremely small partial widths, which can be neglected compared to all other decay channels.}
\label{tab:Lambdab1F}
\renewcommand\arraystretch{1.15}
\begin{tabular*}{86mm}{@{\extracolsep{\fill}}lccc}
\toprule[1.00pt]
\toprule[1.00pt]
Decay channels             &$M_f$ (MeV) &$\Lambda_b(1F,5/2^-)$ &$\Lambda_b(1F,7/2^-)$ \\
\midrule[0.75pt]
$\Sigma_b(1S,1/2^+)\pi$    &5816        & 0.6                  & 0.3                  \\
$\Sigma_b(1S,3/2^+)\pi$    &5835        & 0.6                  & 1.0                  \\
$\Sigma_{b2}(1P,3/2^-)\pi$ &6082        &11.7                  & 0.2                  \\
$\Sigma_{b2}(1P,5/2^-)\pi$ &6089        & 1.0                  &12.3                  \\
$N \bar{B}$                &            &24.9                  & 5.9                  \\
$N \bar{B}^*$              &            &20.2                  &43.9                  \\
$\cdots$                   &            & 0.4                  & 0.4                  \\
\midrule[0.75pt]
Total                      &            &59.4                  &64.0                  \\
\bottomrule[1.00pt]
\bottomrule[1.00pt]
\end{tabular*}
\end{table}

\subsection{$\Xi_b(1F)$ baryons}
As with the $\Lambda_b$ states, significant progress has been made in understanding the $\Xi_b(1S)$, $\Xi_b(1P)$, and $\Xi_b(1D)$ states~\cite{DELPHI:1995jet,LHCb:2023zpu,LHCb:2021ssn}. However, the $\Xi_b(1F)$ state remains largely unexplored. To encourage further experimental investigation, we provide a comprehensive study of their strong decay characteristics. The numerical results are presented in Table~\ref{tab:Xib1F}. According to Table \ref{tab:Xib1F}, the $\Xi_b(1F,5/2^-)$ and $\Xi_b(1F,7/2^-)$ states exhibit total decay widths of approximately 39.2 MeV and 19.9 MeV, respectively, indicating their relatively narrow nature. The $\Sigma\bar{B}$ decay channel makes up 68.9$\%$ of the total decay width of the $\Xi_b(1F,5/2^-)$ state, which translates to a partial width of 27.0 MeV. This dominance suggests that the $\Sigma\bar{B}$ channel is the key decay channel for the experimental verification of the $\Xi_b(1F,5/2^-)$ state. Furthermore, the $\Xi_b(1F,7/2^-)$ state predominantly decays via the $\Lambda\bar{B}^*$ and $\Sigma\bar{B}^*$ channels, with branching fractions of 30.2$\%$ and 27.1$\%$, respectively. The relatively small decay widths of other channels highlight the importance of these two channels for the detection of the $\Xi_b(1F,7/2^-)$ state. 

These results show that $\Lambda\bar{B}^{(*)}$ and $\Sigma\bar{B}^{(*)}$ channels are important decay modes of $\Xi_b(1F)$ states, and they mainly determine the total decay width of $\Xi_b(1F)$ states. Specifically, the decay processes of $\Xi_b(1F, 5/2^-)$ to $\Lambda\bar{B}$ and $\Sigma\bar{B}$ happen via $D$-wave, while $\Xi_b(1F, 7/2^-)$ decays to $\Lambda\bar{B}$ and $\Sigma\bar{B}$ through $G$-wave. Therefore, the decay width of $\Xi_b(1F, 5/2^-)$ into these two channels is larger than that of $\Xi_b(1F, 7/2^-)$ to both.

\begin{table}
\centering
\caption{The calculated partial and total decay widths (in units of MeV) of the $\Xi_b(1F)$ states.
$M_f$ denotes the mass of the outgoing singly bottom baryons. The notation ``$\cdots$" refers to decay channels with negligible partial widths, which are considered insignificant compared to the other decay channels.}
\label{tab:Xib1F}
\renewcommand\arraystretch{1.15}
\begin{tabular*}{86mm}{@{\extracolsep{\fill}}lccc}
\toprule[1.00pt]
\toprule[1.00pt]
Decay channels                 &$M_f$ (MeV) &$\Xi_b(1F,5/2^-)$ &$\Xi_b(1F,7/2^-)$ \\
\midrule[0.75pt]
$\Xi^\prime_{b2}(1P,3/2^-)\pi$ &6211        & 2.0              & 0.1              \\
$\Xi^\prime_{b2}(1P,5/2^-)\pi$ &6220        & 0.2              & 2.2              \\
$\Sigma_b(1S,1/2^+)\bar{K}$    &5816        & 1.4              & 0.5              \\
$\Sigma_b(1S,3/2^+)\bar{K}$    &5835        & 1.1              & 2.2              \\
$\Lambda \bar{B}$              &            & 2.8              & 1.1              \\
$\Sigma \bar{B}$               &            &27.0              & 1.9              \\
$\Lambda \bar{B}^*$            &            & 3.1              & 6.0              \\
$\Sigma \bar{B}^*$             &            & 1.1              & 5.4              \\
$\cdots$                       &            & 0.5              & 0.5              \\
\midrule[0.75pt]
Total                          &            &39.2              &19.9              \\
\bottomrule[1.00pt]
\bottomrule[1.00pt]
\end{tabular*}
\end{table}

\subsection{$\Sigma_b(1F)$ baryons}
The $\Sigma_{b}$ and $\Lambda_b$ states constitute flavor partners, representing flavor-symmetric and flavor-antisymmetric states, respectively. While the $\Lambda_b$ states have received significant attention, the properties of the $\Sigma_{b}$ states remain poorly understood and require further investigation. We further study the strong decay behavior of the $\Sigma_{b}(1F)$ states to achieve a more profound understanding of these states.

Table \ref{tab:Sigmab1F} presents the calculated decay widths including $\Sigma_{b2}(1F,3/2^-)$, $\Sigma_{b2}(1F,5/2^-)$, $\Sigma_{b3}(1F,5/2^-)$, $\Sigma_{b3}(1F,7/2^-)$, $\Sigma_{b4}(1F,7/2^-)$, and $\Sigma_{b4}(1F,9/2^-)$. The results indicate that the decay widths of $\Sigma_{b2}(1F,3/2^-)$, $\Sigma_{b2}(1F,5/2^-)$, $\Sigma_{b3}(1F,5/2^-)$, and $\Sigma_{b3}(1F,7/2^-)$ are approximately 75 MeV, while those of $\Sigma_{b4}(1F,7/2^-)$ and $\Sigma_{b4}(1F,9/2^-)$ are significantly narrower, around 40 MeV. A detailed discussion of the partial decay widths for each state follows.

The $\Sigma_{b2}(1F,3/2^-)$ state exhibits dominant decay modes into $\Lambda_b(1D,3/2^+)\pi$, $\Lambda_b(1F,5/2^-)\pi$, and $\Sigma_{b2}(1D,3/2^+)\pi$, with corresponding branching fractions of 14.0$\%$, 21.4$\%$, and 22.5$\%$, respectively. Similarly, the $\Sigma_{b2}(1F,5/2^-)$ state preferentially decays into $\Lambda_b(1D,5/2^+)\pi$, $\Lambda_b(1F,7/2^-)\pi$, and $\Sigma_{b2}(1D,5/2^+)\pi$, with branching fractions of 14.3$\%$, 20.3$\%$, and 23.8$\%$, respectively. In the case of $\Sigma_{b3}(1F,5/2^-)$, the largest branching fractions are associated with the $\Lambda_b(1D,3/2^+)\pi$ and $\Sigma_{b3}(1D,5/2^+)\pi$ channels, with values of 19.5$\%$ and 34.0$\%$, respectively. The $\Sigma_{b3}(1F,7/2^-)$ state is found to decay predominantly into $\Lambda_b(1D,5/2^+)\pi$ and $\Sigma_{b3}(1D,7/2^+)\pi$, with branching fractions of 21.8$\%$ and 33.4$\%$, respectively. For $\Sigma_{b4}(1F,7/2^-)$, several decay channels exhibit branching fractions in the range of 11$\%$ to 18$\%$, including $\Lambda_b(1P,1/2^-)\pi$, $\Lambda_b(1P,3/2^-)\pi$, $\Lambda_b(1D,3/2^+)\pi$, $\Sigma_{b1}(1P,3/2^-)\pi$, and $N \bar{B}^*$ channels. The $\Sigma_{b4}(1F,9/2^-)$ state primarily decays through the $\Lambda_b(1P,3/2^-)\pi$, $\Lambda_b(1D,5/2^+)\pi$, and $\Sigma_{b2}(1P,5/2^-)\pi$ channels, which account for 20.7$\%$, 12.4$\%$, and 15.2$\%$ of the total width, respectively. The widths of the remaining decay channels are relatively small. 

Some physical reasons lead to the occurrence of the above results. If the strong decay channel includes a singly heavy baryon and a light flavor meson, the quantum numbers of the light degrees of freedom must be conserved due to the heavy quark spin symmetry. The violation of light degrees of freedom conservation in the decay processes of $\Sigma_{b2}(1F,3/2^-) \to \Sigma_{b0}(1P,1/2^-)\pi$, $\Sigma_{b2}(1F,5/2^-) \to \Sigma_{b0}(1P,1/2^-)\pi$, $\Sigma_{b3}(1F,5/2^-) \to \Lambda_b(2S,1/2^+)\pi$, $\Sigma_{b3}(1F,7/2^-) \to \Lambda_b(2S,1/2^+)\pi$, $\Sigma_{b4}(1F,7/2^-) \to \Sigma_{b0}(1P,1/2^-)\pi$, and $\Sigma_{b4}(1F,9/2^-) \to \Sigma_{b0}(1P,1/2^-)\pi$ makes these processes forbidden. Because the $\Sigma_{b4}(1F,7/2^-)$ and $\Sigma_{b4}(1F,9/2^-)$ states do not have $S$-wave decay channels, these two baryons have smaller decay widths. It should be emphasized that the processes $\Sigma_{bj_\ell}(1F,J^P)\to\Lambda_{bj_\ell}(1D,J^{-P})\pi$ and $\Sigma_{bj_\ell}(1F,J^P)\to\Sigma_{bj_\ell}(1D,J^{-P})\pi$ occur mainly through $S$-wave decays, which explains why the branching ratios of these processes are relatively large. Here, we notice that these two processes represent the first occurrence of decay channels involving $S$-wave decay in this article.

Based on these results, the $\Lambda_b(1D)\pi$ and $\Sigma_{b2}(1D)\pi$ channels are good candidates for experimental searches for $\Sigma_{b2}(1F,3/2^-)$, $\Sigma_{b2}(1F,5/2^-)$, and $\Sigma_{b3}(1F,7/2^-)$. Given that $\Lambda_b(1D)$ and $\Sigma_{b2}(1D)$ are predicted to decay into $\Sigma_b\pi$ and $\Lambda_b\pi$, respectively, according to other theoretical studies \cite{Yao:2018jmc}, we propose searching for these $\Sigma_b(1F)$ states via the $\Sigma_b\pi\pi$ and $\Lambda_b\pi\pi$ channels. Moreover, the branching ratio information suggests that the $\Sigma_b(1P)\pi$ and $\Lambda_b(1P)\pi$ channels are also viable candidates for searching for $\Sigma_{b}(1F,7/2^-)$ and $\Sigma_{b}(1F,9/2^-)$.

\begin{table*}
\centering
\caption{The calculated partial and total decay widths (in units of MeV) of the $\Sigma_b(1F)$ states. $M_f$ is the mass of the final singly bottom. The ellipsis  ``$\cdots$" denotes decay channels with extremely small partial widths, which can be neglected compared to all other decay channels. The number ``$0.0$" indicates that the corresponding partial width is smaller than $0.1$ MeV. The symbol ``$\times$" signifies that the initial and final states of the corresponding decay process cannot couple. The mark ``$-$" indicates that the initial state lies below the threshold of a specific decay channel.}
\label{tab:Sigmab1F}
\renewcommand\arraystretch{1.15}
\begin{tabular*}{178mm}{@{\extracolsep{\fill}}lccccccc}
\toprule[1.00pt]
\toprule[1.00pt]
Decay channels             &$M_f$ (MeV) &$\Sigma_{b2}(1F,3/2^-)$ &$\Sigma_{b2}(1F,5/2^-)$ &$\Sigma_{b3}(1F,5/2^-)$ &$\Sigma_{b3}(1F,7/2^-)$ &$\Sigma_{b4}(1F,7/2^-)$ &$\Sigma_{b4}(1F,9/2^-)$ \\
\midrule[0.75pt]
$\Lambda_b(2S,1/2^+)\pi$   &6072        & 7.1                    & 7.5                    &$\times$                &$\times$                & 5.1                    & 5.2                    \\
$\Lambda_b(1P,1/2^-)\pi$   &5912        & 2.7                    & 0.1                    & 1.6                    & 0.9                    & 5.5                    & 2.1                    \\
$\Lambda_b(1P,3/2^-)\pi$   &5920        & 0.6                    & 3.2                    & 2.0                    & 2.7                    & 5.0                    & 8.7                    \\
$\Lambda_b(2P,1/2^-)\pi$   &6290        & 3.6                    & 0.0                    & 0.0                    & 0.0                    & 0.0                    & 0.0                    \\
$\Lambda_b(2P,3/2^-)\pi$   &6298        & 0.6                    & 3.9                    & 0.0                    & 0.1                    & 0.0                    & 0.0                    \\
$\Lambda_b(1D,3/2^+)\pi$   &6146        &10.9                    & 0.6                    &13.9                    & 2.8                    & 4.6                    & 0.4                    \\
$\Lambda_b(1D,5/2^+)\pi$   &6152        & 1.0                    &11.2                    & 4.5                    &15.9                    & 0.9                    & 5.2                    \\
$\Lambda_b(1F,5/2^-)\pi$   &6344        &16.7                    & 0.9                    & 2.2                    & 0.1                    &$-$                     &$-$                     \\
$\Lambda_b(1F,7/2^-)\pi$   &6348        & 0.0                    &15.9                    & 0.1                    & 2.3                    &$-$                     &$-$                     \\
$\Sigma_{b0}(1P,1/2^-)\pi$ &6097        &$\times$                &$\times$                & 2.4                    & 2.5                    &$\times$                &$\times$                \\
$\Sigma_{b1}(1P,1/2^-)\pi$ &6086        & 0.1                    & 0.9                    & 4.0                    & 2.3                    & 1.8                    & 0.1                    \\
$\Sigma_{b1}(1P,3/2^-)\pi$ &6098        & 1.4                    & 0.9                    & 4.7                    & 6.4                    & 0.7                    & 2.2                    \\
$\Sigma_{b2}(1P,3/2^-)\pi$ &6082        & 2.1                    & 2.1                    & 1.6                    & 0.9                    & 5.8                    & 1.1                    \\
$\Sigma_{b2}(1P,5/2^-)\pi$ &6089        & 3.1                    & 3.2                    & 1.2                    & 2.2                    & 1.9                    & 6.4                    \\
$\Sigma_{b2}(1D,3/2^+)\pi$ &6313        &17.6                    & 1.4                    & 0.6                    & 0.1                    & 0.0                    & 0.0                    \\
$\Sigma_{b2}(1D,5/2^+)\pi$ &6319        & 1.5                    &18.6                    & 0.1                    & 0.7                    & 0.0                    & 0.0                    \\
$\Sigma_{b3}(1D,5/2^+)\pi$ &6286        & 4.2                    & 0.9                    &24.3                    & 0.4                    & 0.7                    & 0.1                    \\
$\Sigma_{b3}(1D,7/2^+)\pi$ &6290        & 0.8                    & 4.2                    & 0.4                    &24.4                    & 0.1                    & 0.7                    \\
$N \bar{B}$                &            & 0.2                    & 0.0                    & 0.3                    & 2.0                    & 0.1                    & 3.9                    \\
$\Delta \bar{B}$           &            & 2.1                    & 0.5                    &$-$                     &$-$                     &$-$                     &$-$                     \\
$N \bar{B}^*$              &            & 0.4                    & 0.9                    & 5.0                    & 3.8                    & 7.5                    & 4.3                    \\
$\cdots$                   &            & 1.4                    & 1.4                    & 2.5                    & 2.6                    & 1.6                    & 1.6                    \\
\midrule[0.75pt]
Total                      &            &78.1                    &78.3                    &71.4                    &73.1                    &41.3                    &42.0                    \\
\bottomrule[1.00pt]
\bottomrule[1.00pt]
\end{tabular*}
\end{table*}

\subsection{$\Xi_b^{\prime}(1F)$ baryons}
The $\Xi_b^{\prime}$ baryons are the flavor partners of the $\Xi_b$ baryons, but their understanding is less complete compared to the latter. To enhance the understanding of the $\Xi_b^{\prime}$ states, we have conducted a further study on the strong decays of the $\Xi_b^{\prime}(1F)$ states, and present our results in Table \ref{tab:Xibp1F}.

Our results indicate that the total widths of $\Xi^\prime_{b2}(1F,3/2^-)$, $\Xi^\prime_{b2}(1F,5/2^-)$, $\Xi^\prime_{b3}(1F,5/2^-)$, and $\Xi^\prime_{b3}(1F,7/2^-)$ are around 86 MeV, while the widths of $\Xi^\prime_{b4}(1F,7/2^-)$ and $\Xi^\prime_{b4}(1F,9/2^-)$ are approximately 40 MeV. We find that the $\Xi^\prime_{b2}(1F,3/2^-)$ state has two dominant decay channels: $\Lambda_b(1D,3/2^+)\bar{K}$ and $\Sigma_{b1}(1P,1/2^-)\bar{K}$, which contribute 36.7$\%$ and 18.6$\%$ to the total width, respectively. For the $\Xi^\prime_{b2}(1F,5/2^-)$ state, the $\Lambda_b(1D,5/2^+)\bar{K}$ and $\Sigma_{b1}(1P,3/2^-)\bar{K}$ channels have particularly large branching ratios, contributing 35.0$\%$ and 21.6$\%$ to the total width, respectively. The $\Sigma_{b2}(1P,3/2^-)\bar{K}$ channel is the primary decay mode of $\Xi^\prime_{b3}(1F,5/2^-)$, accounting for 41.2$\%$ of its total width. For the $\Xi^\prime_{b3}(1F,7/2^-)$ decay widths, the $\Sigma_{b2}(1P,5/2^-)\bar{K}$ channel is dominant, accounting for 41.2$\%$ of the total width. The $\Lambda_b(1P,1/2^-)\bar{K}$ and $\Lambda \bar{B}^*$ channels are two important decay modes of $\Xi^\prime_{b4}(1F,7/2^-)$, contributing 14.4$\%$ and 21.1$\%$ to the total width, respectively. For the $\Xi^\prime_{b4}(1F,9/2^-)$ decay pattern, the $\Xi_b(1P,3/2^-)\pi$, $\Lambda_b(1P,3/2^-)\bar{K}$, $\Lambda \bar{B}$, $\Sigma \bar{B}$, and $\Lambda \bar{B}^*$ channels are identified as several good decay modes, with their partial widths accounting for approximately 10-20$\%$ of the total width.

We should emphasize that the $\Xi_{bj_\ell}^{\prime}(1F,J^P) \to \Xi_{bj_\ell}(1D,J^{-P})\pi$, $\Xi_{bj_\ell}^{\prime}(1F,J^P) \to \Xi_{bj_\ell}^{\prime}(1D,J^{-P})\pi$, and $\Xi_{bj_\ell}^{\prime}(1F,J^P) \to \Lambda_{bj_\ell}(1D,J^{-P})\bar{K}$ processes all predominantly decay through $S$-wave decays, resulting in relatively large branching ratios for these decay processes, where ``$-P$" implies that the parity is reversed relative to the initial state. The process of $\Xi_{bj_\ell}^{\prime}(1F,J^P) \to \Sigma_{bj_\ell}(1D,J^{-P}) \bar{K}$ also decays via the $S$-wave, but the initial state mass of this process is below the threshold of the final $\Sigma_{bj_\ell}(1D,J^{-P}) \bar{K}$ channel, making this process kinematically forbidden. Moreover, the process $\Xi_{bj_\ell}^{\prime}(1F,J^P)\to\Sigma_{bj_{\ell-1}}(1P,(J-1)^{P})\pi$ proceeds including a $P$-wave decay, which also contributes significantly to the total decay width. Apart from these channels, the decay modes involving a $1D$-wave singly bottom baryon ($\Lambda_b(1D)$, $\Xi_b(1D)$, and $\Xi_b^{\prime}(1D)$) with a pseudoscalar meson ($\pi$ and $\bar{K}$) can only happen through higher partial waves, so the partial widths of these processes may contribute particularly little to the total width. Because the $\Xi^{\prime}_{b4}(1F,7/2^-)$ and $\Xi^{\prime}_{b4}(1F,9/2^-)$ states do not have decay channels via $S$-wave under heavy quark symmetry, the total widths of the two states are relatively narrow. Additionally, the decay processes of $\Sigma_b(1F)$ to $\Lambda_b(1S)\pi$ and $\Sigma_b(1S)\pi$ channels can only happen through higher partial waves, which causes the contributions from these decay channels to be particularly small.

These results suggest that we should consider the significant role of the $\Lambda_b(1D)\bar{K}$ channel in the observation of the $\Xi^\prime_{b2}(1F,3/2^-)$ and $\Xi^\prime_{b2}(1F,5/2^-)$ states, as well as the important role of the $\Sigma_{b}(1P)\bar{K}$ channel in the search for the $\Xi^\prime_{b2}(1F,3/2^-)$, $\Xi^\prime_{b2}(1F,5/2^-)$, $\Xi^\prime_{b3}(1F,5/2^-)$, and $\Xi^\prime_{b3}(1F,7/2^-)$ states.
\begin{table*}
	\centering
	\caption{The calculated partial and total decay widths (in units of MeV) of the $\Xi_b^{\prime}(1F)$ states. All markings are consistent with those in Table~\ref{tab:Sigmab1F}.}
	\label{tab:Xibp1F}
	\renewcommand\arraystretch{1.15}
\begin{tabular*}{178mm}{@{\extracolsep{\fill}}lccccccc}
	\toprule[1.00pt]
	\toprule[1.00pt]
Decay channels                 &$M_f$ (MeV) &$\Xi^\prime_{b2}(1F,3/2^-)$ &$\Xi^\prime_{b2}(1F,5/2^-)$ &$\Xi^\prime_{b3}(1F,5/2^-)$ &$\Xi^\prime_{b3}(1F,7/2^-)$ &$\Xi^\prime_{b4}(1F,7/2^-)$ &$\Xi^\prime_{b4}(1F,9/2^-)$ \\
\midrule[0.75pt]
$\Xi_b(2S,1/2^+)\pi$           &6253        & 0.5                        & 0.6                        &$\times$                    &$\times$                    & 1.1                        & 1.2                        \\
$\Xi_b(1P,1/2^-)\pi$           &6087        & 0.8                        & 0.0                        & 0.7                        & 0.4                        & 3.8                        & 0.4                        \\
$\Xi_b(1P,3/2^-)\pi$           &6095        & 0.2                        & 0.9                        & 0.8                        & 1.2                        & 1.8                        & 5.2                        \\
$\Xi_b(1D,3/2^+)\pi$           &6327        & 3.2                        & 0.7                        & 5.2                        & 0.8                        & 3.4                        & 0.0                        \\
$\Xi_b(1D,5/2^+)\pi$           &6333        & 1.0                        & 3.4                        & 1.3                        & 5.7                        & 0.4                        & 3.6                        \\
$\Xi^\prime_{b1}(1P,3/2^-)\pi$ &6220        & 0.3                        & 0.3                        & 1.1                        & 1.5                        & 0.2                        & 0.8                        \\
$\Xi^\prime_{b2}(1P,3/2^-)\pi$ &6211        & 0.4                        & 0.4                        & 0.6                        & 0.1                        & 1.5                        & 0.3                        \\
$\Xi^\prime_{b2}(1P,5/2^-)\pi$ &6220        & 0.5                        & 0.6                        & 0.2                        & 0.8                        & 0.5                        & 1.6                        \\
$\Xi^\prime_{b2}(1D,3/2^+)\pi$ &6440        & 6.0                        & 0.2                        & 0.2                        & 0.0                        & 0.0                        & 0.0                        \\
$\Xi^\prime_{b2}(1D,5/2^+)\pi$ &6446        & 0.2                        & 6.1                        & 0.0                        & 0.2                        & 0.0                        & 0.0                        \\
$\Xi^\prime_{b3}(1D,5/2^+)\pi$ &6424        & 0.6                        & 0.1                        & 8.5                        & 0.1                        & 0.2                        & 0.0                        \\
$\Xi^\prime_{b3}(1D,7/2^+)\pi$ &6428        & 0.1                        & 0.6                        & 0.1                        & 8.6                        & 0.0                        & 0.2                        \\
$\Lambda_b(1S,1/2^+)\bar{K}$   &5619        & 0.0                        & 0.0                        &$\times$                    &$\times$                    & 1.2                        & 1.2                        \\
$\Lambda_b(1P,1/2^-)\bar{K}$   &5912        & 0.9                        & 0.2                        & 1.9                        & 1.1                        & 5.7                        & 1.2                        \\
$\Lambda_b(1P,3/2^-)\bar{K}$   &5920        & 0.7                        & 1.2                        & 2.2                        & 3.1                        & 3.4                        & 8.1                        \\
$\Lambda_b(1D,3/2^+)\bar{K}$   &6146        &31.6                        & 0.2                        &$-$                         &$-$                         &$-$                         &$-$                         \\
$\Lambda_b(1D,5/2^+)\bar{K}$   &6152        & 0.1                        &32.1                        &$-$                         &$-$                         &$-$                         &$-$                         \\
$\Sigma_b(1S,1/2^+)\bar{K}$    &5816        & 0.3                        & 0.1                        & 1.0                        & 1.4                        & 1.0                        & 0.7                        \\
$\Sigma_b(1S,3/2^+)\bar{K}$    &5835        & 0.4                        & 0.6                        & 2.7                        & 2.5                        & 1.4                        & 1.8                        \\
$\Sigma_{b1}(1P,1/2^-)\bar{K}$ &6086        &16.0                        & 0.1                        & 0.8                        & 0.5                        & 0.0                        & 0.0                        \\
$\Sigma_{b1}(1P,3/2^-)\bar{K}$ &6098        & 3.4                        &19.8                        & 0.5                        & 0.8                        & 0.0                        & 0.0                        \\
$\Sigma_{b2}(1P,3/2^-)\bar{K}$ &6082        & 6.7                        & 1.7                        &35.2                        & 0.0                        & 0.2                        & 0.0                        \\
$\Sigma_{b2}(1P,5/2^-)\bar{K}$ &6089        & 2.0                        & 7.5                        & 2.4                        &37.3                        & 0.0                        & 0.1                        \\
$\Lambda_b(1S,1/2^+)\bar{K}^*$ &5619        & 0.6                        & 0.6                        & 1.3                        & 1.3                        & 0.2                        & 0.2                        \\
$\Lambda \bar{B}$              &            & 0.5                        & 0.0                        & 0.6                        & 2.3                        & 0.1                        & 4.8                        \\
$\Sigma \bar{B}$               &            & 2.8                        & 0.1                        & 2.5                        & 2.7                        & 0.1                        & 4.4                        \\
$\Lambda \bar{B}^*$            &            & 1.2                        & 2.6                        & 5.9                        & 5.4                        & 8.3                        & 4.8                        \\
$\Sigma \bar{B}^*$             &            & 4.6                        &10.5                        & 7.2                        &10.5                        & 3.9                        & 2.4                        \\
$\cdots$                       &            & 0.5                        & 0.6                        & 2.5                        & 2.2                        & 1.1                        & 0.5                        \\
\midrule[0.75pt]
Total                          &            &86.1                        &91.8                        &85.4                        &90.5                        &39.5                        &43.5                        \\
	\bottomrule[1.00pt]
	\bottomrule[1.00pt]
\end{tabular*}
\end{table*}

\subsection{$\Omega_b(1F)$ baryons}
Table \ref{tab:Omegab1F} presents our calculated strong decay widths for the $\Omega_b(1F)$ states. The results indicate that $\Omega_{b2}(1F,3/2^-)$, $\Omega_{b2}(1F,5/2^-)$, $\Omega_{b3}(1F,5/2^-)$, and $\Omega_{b3}(1F,7/2^-)$ are relatively broad states with total widths of 151.2, 189.9, 141.6, and 175.4 MeV, respectively. 
The $\Omega_{b4}(1F,7/2^-)$ and $\Omega_{b4}(1F,9/2^-)$ have relatively small decay widths of 63.2 and 89.0 MeV, respectively. In the $\Omega_{b2}(1F,3/2^-)$ state, the $\Xi \bar{B}$ and $\Xi \bar{B}^*$ channels have significant branching fractions, accounting for 36.9$\%$ and 45.9$\%$ of the total width, respectively. Among all decay channels of $\Omega_{b2}(1F,5/2^-)$, $\Xi \bar{B}$ is the dominant channel, with its partial width accounting for 85.7$\%$ of the total width. The $\Xi \bar{B}$ and $\Xi \bar{B}^*$ channels are also two important decay channels of $\Omega_{b3}(1F,5/2^-)$, with their partial widths accounting for 26.1$\%$ and 51.1$\%$ of the total width, respectively. The $\Xi \bar{B}^*$ channel is the most prominent decay channel of $\Omega_{b3}(1F,7/2^-)$, with its partial width accounting for 69.4$\%$ of the total width. Additionally, $\Xi^\prime_{b2}(1P,3/2^-)\bar{K}$ and $\Xi^\prime_{b2}(1P,5/2^-)\bar{K}$ are non-negligible channels, with their partial widths accounting for approximately 12$\%$ of the total width. In the decay channels of $\Omega_{b4}(1F,7/2^-)$, the $\Xi \bar{B}^*$ channel is also the dominant decay channel, accounting for 68.4$\%$ of the total width, and $\Xi_b(1P,1/2^-)\bar{K}$ is a slightly prominent channel, accounting for 18.7$\%$ of the total width. For $\Omega_{b4}(1F,9/2^-)$, $\Xi \bar{B}$ and $\Xi \bar{B}^*$ are relatively important decay channels, with their partial widths accounting for 49.3$\%$ and 29.6$\%$ of the total width, respectively.

In the singly bottom baryon $\Omega_b$, the heavy quark $b$ behaves as a spectator, while the two strange quarks form a $ss$ cluster. Among all its decay channels, the nature of $\Xi^{(*)}\bar{B}^{(*)}$ channels can correspond to this property of the $\Omega_b(1F)$ states, and the $\rho$-mode is well preserved in the decay process, suggesting that the $\Xi^{(*)}\bar{B}^{(*)}$ channels should have a considerable branching fractions. Our results show that the $\Xi\bar{B}^{(*)}$ channels indeed have a sizable partial width. Moreover, the $\Xi^* \bar{B}$ and $\Xi^*\bar{B}^*$ channels have thresholds above the mass range of the $\Omega_b(1F)$ states, implying that their contributions to the $\Omega_b(1F)$ are negligible. We also notice that all six $\Omega_b(1F)$ baryons do not decay through the decay channels involving $S$-wave decay due to the thresholds for the $\Xi_b(1D)\bar{K}$ and $\Xi_b^{\prime}(1D)\bar{K}$ channels with $S$-wave decay exceed the masses of these $\Omega_b(1F)$ baryons. In addition, the processes $\Omega_{b4}(1F,J^P)\to \Xi^{\prime}_{b{j_l}}(1P)\bar{K}$ decay through partial waves higher than $P$-wave. This suppression results in partial widths that are significantly smaller than those of the other four $\Omega_b(1F)$ states. The smaller partial widths are one reason why the total decay widths of these two $\Omega_{b4}$ states are narrower than those of the other four $\Omega_b(1F)$ states.

These results suggest that the $\Xi \bar{B}$ channel plays a considerable role in searching for the $\Omega_{b2}(1F,3/2^-)$, $\Omega_{b3}(1F,5/2^-)$, $\Omega_{b3}(1F,7/2^-)$, and $\Omega_{b4}(1F,9/2^-)$ states. Furthermore, the $\Xi \bar{B}^*$ channel is crucial for the discovery of the $\Omega_b(1F)$ states. In addition, we should not overlook the contribution of $\Xi^\prime_{b2}(1P)\bar{K}$ to the total widths of $\Omega_{b2}(1F,3/2^-)$, $\Omega_{b2}(1F,5/2^-)$, $\Omega_{b3}(1F,5/2^-)$, and $\Omega_{b3}(1F,7/2^-)$, and the non-negligible role of the $\Xi_b(1P,1/2^-)\bar{K}$ channel in the discovery of $\Omega_{b4}(1F,7/2^-)$ and $\Omega_{b4}(1F,9/2^-)$.

\begin{table*}
	\centering
	\caption{The calculated partial and total decay widths (in units of MeV) of the $\Omega_b(1F)$ states. All markings are identical to those in Table~\ref{tab:Sigmab1F}.}
	\label{tab:Omegab1F}
	\renewcommand\arraystretch{1.15}
	\begin{tabular*}{178mm}{@{\extracolsep{\fill}}lccccccc}
		\toprule[1.00pt]
		\toprule[1.00pt]
		Decay channels                     &$M_f$ (MeV) &$\Omega_{b2}(1F,3/2^-)$ &$\Omega_{b2}(1F,5/2^-)$ &$\Omega_{b3}(1F,5/2^-)$ &$\Omega_{b3}(1F,7/2^-)$ &$\Omega_{b4}(1F,7/2^-)$ &$\Omega_{b4}(1F,9/2^-)$ \\
\midrule[0.75pt]
$\Xi_b(1S,1/2^+)\bar{K}$           &5795        &  0.0                   &  0.0                   &$\times$                &$\times$                & 1.9                    & 1.9                    \\
$\Xi_b(1P,1/2^-)\bar{K}$           &6087        &  2.0                   &  1.3                   &  2.3                   &  1.4                   &11.8                    & 0.6                    \\
$\Xi_b(1P,3/2^-)\bar{K}$           &6095        &  3.0                   &  3.9                   &  2.6                   &  3.7                   & 4.2                    &15.0                    \\
$\Xi^\prime_b(1S,3/2^+)\bar{K}$    &5950        &  0.3                   &  0.4                   &  1.2                   &  1.3                   & 0.7                    & 0.9                    \\
$\Xi^\prime_{b1}(1P,1/2^-)\bar{K}$ &6206        & 12.2                   &  0.0                   &  0.4                   &  0.3                   & 0.1                    & 0.0                    \\
$\Xi^\prime_{b1}(1P,3/2^-)\bar{K}$ &6220        &  2.2                   & 13.8                   &  0.2                   &  0.3                   & 0.0                    & 0.0                    \\
$\Xi^\prime_{b2}(1P,3/2^-)\bar{K}$ &6211        &  4.5                   &  0.6                   & 22.6                   &  0.0                   & 0.1                    & 0.0                    \\
$\Xi^\prime_{b2}(1P,5/2^-)\bar{K}$ &6220        &  0.6                   &  4.5                   &  1.4                   & 22.3                   & 0.0                    & 0.1                    \\
$\Xi \bar{B}$                      &            & 55.8                   &  1.4                   & 36.9                   & 22.7                   & 0.7                    &43.9                    \\
$\Xi \bar{B}^*$                    &            & 69.4                   &162.8                   & 72.4                   &121.7                   &43.2                    &26.3                    \\
$\cdots$                           &            &  1.2                   &  1.2                   &  1.6                   &  1.7                   & 0.5                    & 0.3                    \\
\midrule[0.75pt]
Total                              &            &151.2                   &189.9                   &141.6                   &175.4                   &63.2                    &89.0                    \\
		\bottomrule[1.00pt]
		\bottomrule[1.00pt]
	\end{tabular*}
\end{table*}

\section{Summary}\label{sec:Summary}
This work focuses on the spectroscopic analysis of the experimentally unobserved $1F$-wave singly bottom baryons. We employ the nonrelativistic potential model, combined with the Gaussian expansion method, to calculate their mass spectra. Meanwhile, we also calculate the OZI-allowed two-body strong decay widths with the QPC model to provide crucial insights into their decay behaviors and guide experimental searches.

We analyzed the total and partial decay widths of $1F$-wave singly bottom baryons and identified several noteworthy observations. Firstly, the states of $\Xi_b(1F,5/2^-)$, $\Xi_b(1F,7/2^-)$, $\Sigma_b(1F,9/2^-)$, etc., exhibit relatively narrow widths. These narrow states provide a beneficial scenario for experimental detection. Secondly, concerning the partial decay widths, we have derived the following conclusions. Our calculations highlight the $\Sigma_b(1P)\pi$ and $N\bar{B}^*$ channels as the most promising for identifying the $\Lambda_b(1F)$ states. The $\Sigma\bar{B}$ channel is also significant in the search for the $\Xi_b(1F)$ states. For the $\Sigma_b(1F)$ states, the $\Lambda_b(1D)\pi$ and $\Sigma_b(1D)\pi$ channels emerge as favorable options. For the $\Xi_b^{\prime}(1F)$ states, the $\Lambda_b(1D)\bar{K}$ and $\Sigma_b(1P)\bar{K}$ channels warrant particular attention. The $\Omega_b(1F)$ states exhibits prominent partial widths in the $\Xi^\prime_{b1}(1P)\bar{K}$, $\Xi \bar{B}$, and $\Xi \bar{B}^*$ channels, making them key targets in experimental searches. In addition, decay channels involving a $1S$-wave singly bottom baryon and a pseudoscalar meson are found to have negligible contributions to the total width of all $1F$ wave singly bottom baryons.

Having predicted the total and partial decay widths of these $1F$-wave singly bottom baryons, we have identified the main decay channels for their investigation. With ongoing progress in high-energy experiments, collaborations like LHCb and CMS are expected to detect these states in the near future. We hope our results can assist experimental searches and contribute to the growing understanding of the spectroscopy of singly heavy baryons.

\section*{ACKNOWLEDGMENTS}
We are very grateful to Prof. Zhan-Wei Liu for helpful discussions. This project is partially supported by the National Natural Science Foundation of China under Grants No. 12405098, No. 12335001, No. 12247101, the Talent Scientific Fund of Lanzhou University, the Fundamental Research Funds for the Central Universities (Grant No. lzujbky-2024-jdzx06), the Natural Science Foundation of Gansu Province (No. 22JR5RA389, No.25JRRA799), and the ‘111 Center’ under Grant No. B20063.

\end{document}